\documentclass[%
reprint,
superscriptaddress,
twocolumn,
bibnotes,
amsmath,amssymb,
]{revtex4-2}
\usepackage{footmisc}
\usepackage{graphicx}
\usepackage{epstopdf}

\usepackage{dcolumn}
\usepackage{bm}
\usepackage{titlesec}
\titleformat{\section}{\raggedright\fontsize{12.5}{25}\bfseries}{\arabic{section}.}{1em}{}
\pdfoutput=1

\begin{document}

\title{\fontsize{15}{19}\selectfont Ultrafast Evolution of Bulk, Surface and Surface Resonance States in Photoexcited Bi$_{2}$Te$_{3}$}

\author{Hamoon Hedayat}
\email{Current address: Institute of Physics II, University of Cologne, D-50937 Cologne, Germany}
\affiliation{IFN-CNR, Dipartimento di Fisica, Politecnico di Milano, 20133 Milan, Italy}
\affiliation{Dipartimento di Fisica, Politecnico di Milano, 20133 Milan, Italy}
\author{Davide Bugini}
\affiliation{Dipartimento di Fisica, Politecnico di Milano, 20133 Milan, Italy}
\author{Hemian Yi}
\affiliation{National Lab for Superconductivity, Institute of Physics, Chinese Academy of Science, 100190 Beijing, China}
\author{Chaoyu Chen}
\affiliation{National Lab for Superconductivity, Institute of Physics, Chinese Academy of Science, 100190 Beijing, China}
\author{Xingjiang Zhou}
\affiliation{National Lab for Superconductivity, Institute of Physics, Chinese Academy of Science, 100190 Beijing, China}
\author{Giulio Cerullo}
\affiliation{Dipartimento di Fisica, Politecnico di Milano, 20133 Milan, Italy}
\author{Claudia Dallera}
\affiliation{Dipartimento di Fisica, Politecnico di Milano, 20133 Milan, Italy}
\author{Ettore Carpene}
\email{ettore.carpene@polimi.it}
\affiliation{IFN-CNR, Dipartimento di Fisica, Politecnico di Milano, 20133 Milan, Italy}

\begin{abstract}
We use circular dichroism (CD) in time- and angle-resolved photoemission spectroscopy (trARPES) to measure the femtosecond charge dynamics in the topological insulator (TI) Bi$_{2}$Te$_{3}$. We detect clear CD signatures from topological surface states (TSS) and surface resonance (SR) states. In time-resolved measurements, independently from the pump polarization or intensity, the CD shows a dynamics which provides access to the unexplored electronic evolution in unoccupied states of Bi$_{2}$Te$_{3}$. In particular, we are able to disentangle the unpolarized electron dynamics in the bulk states from the spin-textured TSS and SR states on the femtosecond timescale. Our study demonstrates that photoexcitation mainly involves the bulk states and is followed by sub-picosecond transport to the surface. This provides essential details on intra- and interband scatterings in the relaxation process of TSS and SR states. Our results reveal the significant role of SRs in the subtle ultrafast interaction between bulk and surface states in TIs. 
\end{abstract}

\maketitle

\section{Introduction\label{Introduction}}
The increasing quest of efficient ultrafast manipulation of spins for applications to spintronics and quantum information technology has pushed the investigation of sub-picosecond dynamics beyond traditional materials exhibiting spin order \cite{kirilyuk,mentink,carpene2015ultrafast,kampfrath}. Three dimensional topological insulators (TIs) have been the subject of such studies due to their conductive topological surface state (TSS), located within the bulk band gap, that hosts spin-polarized Dirac fermions \cite{moore2010,hsieh2008}. In TIs, the combination of spin-orbit coupling (SOC) and time reversal symmetry results in the helical spin-order of the TSS locked to the electron momentum, which leads to immunity against backscattering events \cite{roushan,alpichshev2010,zhang2009,pauly2012}. Recent observations have revealed that the spin-order of the TSS is not the only asset of TIs in spintronic engineering. The unoccupied part of the Dirac cone \cite{sobota2013,bugini2017,niesner2012} and the induced Rashba splitting may open new routes for innovative devices \cite{zhu2011,bahramy2012,wray2011}. In addition, latest studies have reported the existence of surface resonance (SR) states with preferential spin character \cite{nurmamat2013,cacho2015,jozwiak2016,sanchez2017}. SR could explain the complex interaction between TSS and bulk states, as already observed in the mixing of TSS with bulk bands (BBs) \cite{zhang2009,ishida2011,seibel2015}. However, previous studies did not distinguish between the sub-picosecond electron dynamics of SRs and of the nearby BBs \cite{hajlaoui2012,sanchez2017,jozwiak2016,sanchez2016}.

When a TI is optically perturbed, excited electrons decay through normally unoccupied states, including the portion of the Dirac cone above the Fermi level which is spin polarized. Thus, the electronic dynamics can be significantly affected by the spin constraints. This has motivated out-of-equilibrium experiments on TIs. In particular, time- and angle-resolved photoemission spectroscopy (trARPES) provides valuable information on electron interactions \cite{reimann2014,sobota2012,hajlaoui2014,wang2012,crepaldi2012,hajlaoui2012,freyse2018impact} as it gives access to the time evolution of excited carriers in the reciprocal space \cite{hedayat2020non,hedayat2019excitonic,hedayat2020investigation}. However, detecting the spin degrees of freedom requires a more sophisticated method \cite{zhou2009,lee2009,cacho2015}. One experimental approach is to employ circular dichroism (CD). CD-ARPES is based on the coupling between the helicity of the incident photons and the (total) angular momentum of electrons. Here, we define CD as the normalized difference of the ARPES intensities measured with probe beams of opposite circular polarization, CD $=(I_{CR}- I_{CL})/(I_{CR}+ I_{CL})$, where the subscripts CR and CL refer to right and left photon helicity, respectively. A proper interpretation of CD-ARPES can disclose relevant information on the spin of image potential states \cite{nakazawa2016}, magnetic doped TIs \cite{Yilmaz},  Berry curvature
in 2D materials \cite{schuler2020local} and on the evolution of surface localization \cite{kondo2017visualizing}. The relation between CD and spin$/$orbital angular momentum (OAM) is highly complex \cite{wang2011, jung2011,mirhosseini2012,park2012,scholz2011,sanchez2016, bahramy2012,hedayat2018femtosecond,sumida2018enhanced,xu2015photoemission} and a number of investigations reported that the CD signal might be affected by other factors, e.g. the final state effect and the experimental geometry \cite{mulazzi2009,zhu2013,sanchez2014}. Therefore, the definitive link between spin-orbital texture and CD requires a comprehensive analysis and it is beyond the purpose of this work.

\begin{figure*}
\centering
\includegraphics[width=0.80\textwidth]{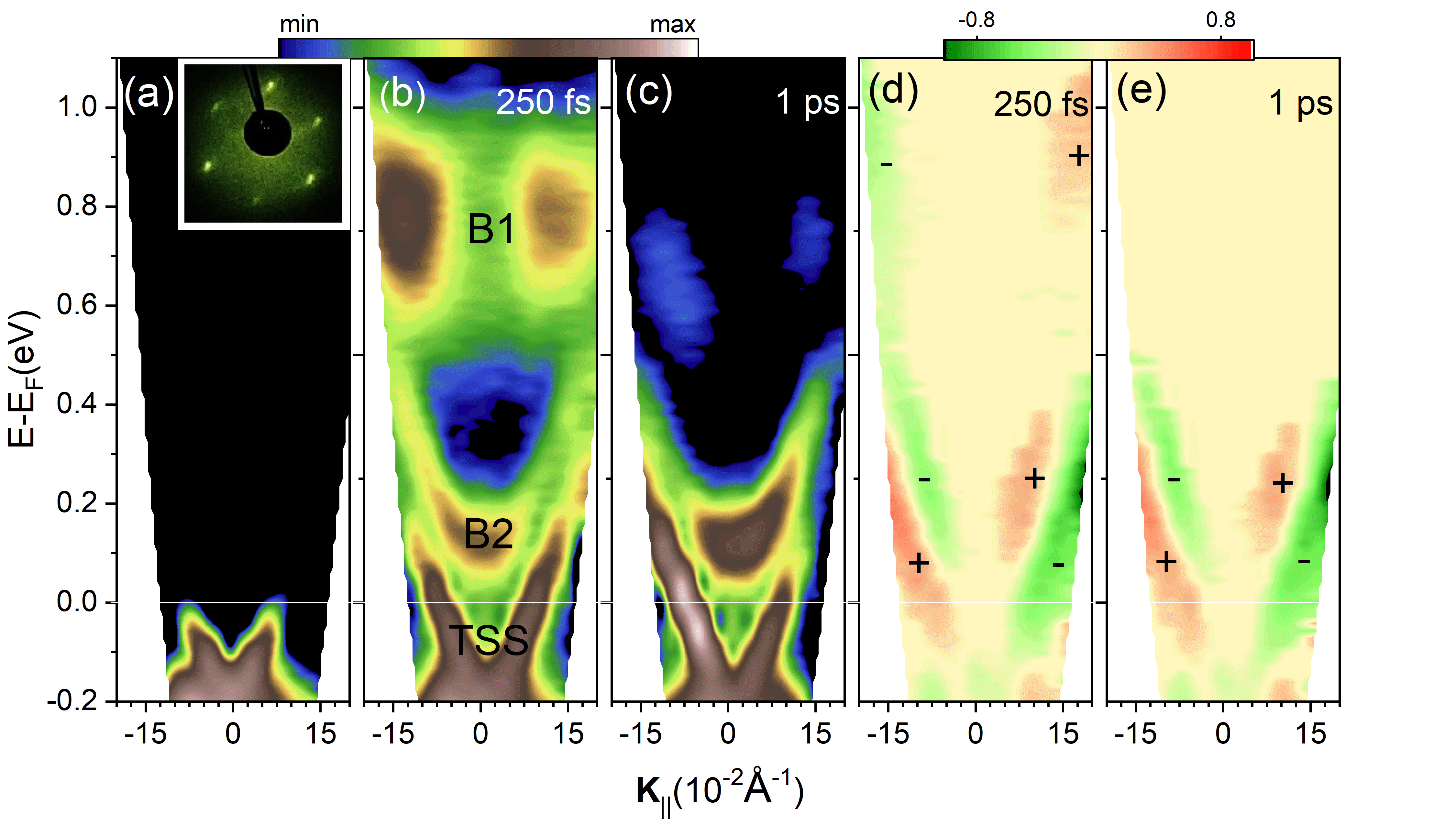}
\caption{ (a)-(c) TrARPES maps of Bi$_2$Te$_3$ along the $\bar{\Gamma}\bar{K}$ for selected pump-probe delays: (a) Negative (unperturbed), (b) $+0.25$~ps, (c) $+1$~ps. Inset (a) shows the LEED pattern. Panels (d) and (e) show the circular dichroism (CD) corresponding the same delays of panels (b) and (c), respectively.}
\label{Fig1}
\end{figure*} 

Here, we exploit CD-trARPES to disentangle bulk and TSS photocurrents. The separation of bulk, TSS and SR dynamics in the trARPES signal is required in order to explain the experimental observations. We present CD-trARPES of the nonequilibrium electron and spin states in Bi$_2$Te$_3$ where only the TSS crosses the Fermi level and the insulating bulk restricts the number of relaxation channels. Our results demonstrate femtosecond decay of unpolarized BBs to TSSs and SRs followed by an electronic accumulation in TSS lasting several ps. The same approach can be generalized to study the bulk and surface electron dynamics in a wide class of TIs and open a route towards their advanced engineering for innovative opto-electronic and spintronics applications.

\section{Results and discussion\label{Results}}
Figures~\ref{Fig1}a-c show the trARPES measurements of a freshly cleaved Bi$_2$Te$_3$ sample recorded along the $\bar{\Gamma}\bar{K}$ direction of the Surface Brillouin Zone (SBZ) in equilibrium (a), 250~fs (b) and 1~ps (c) after excitation by linearly p-polarized pump pulses with 1.85 eV photon energy. The low energy electron diffraction (LEED) pattern of the sample, reported in the upper inset of Fig.~\ref{Fig1}a, confirmed the high quality of the cleaved surface and the absence of surface reconstruction. The Dirac-cone is clearly visible in all three panels. At 250 fs delay (Fig.~\ref{Fig1}b), two unoccupied  bands (B1 and B2) can be seen: one is located 0.8~eV above the Fermi level, the other at lower energy has parabolic dispersion and is well-separated from the TSS. These unoccupied states have been widely investigated in TIs, and predicted theoretically \cite{hajlaoui2012,zhou2015few}. However, their nature is still under debate: while some studies considered them as bulk states \cite{hajlaoui2012,sanchez2016}, other investigations suggested that these bands are SRs \cite{sanchez2017}. We will demonstrate that the photoemission (PE) signal is detected from both SRs and BBs, therefore, B1 and B2 bands in Figs.~\ref{Fig1}b,c are the superposition of BBs and SRs. At 1~ps delay, electrons in the higher energy B1 states have almost completely relaxed, while the two bands at lower energy (B2 and TSS) are still populated, in line with the electronic dynamics of Bi$_2$Te$_3$ previously reported \cite{hajlaoui2012,sanchez2017}.
Figures~\ref{Fig1}d,e show the non-equilibrium CD measured at the same delays as in panels b and c obtained as the difference between the trARPES maps measured with CR and CL light. The dichroic signal is present in all three excited bands, revealing the following features: (i) a nearly perfect anti-symmetric CD behavior with respect to the $\bar{\Gamma}$ point, in agreement with the expected spin symmetry of the Dirac cone \cite{basak2011}; (ii) the CD decreases approaching the center of the SBZ, consistent with the previous reports on the spin structure of the TSSs \cite{sanchez2014anisotropic,wang2011,jung2011}; (iii) the opposite sign of CD in B1 and B2 with respect to the one in TSS: the signs of CD for different bands corresponds to the time dependent one-step PE model calculation and spin-resolved ARPES measurements of the Bi$_2$Te$_3$ unoccupied spin structure \cite{sanchez2017}.

\begin{figure*}
\centering
\includegraphics[width=0.85\textwidth]{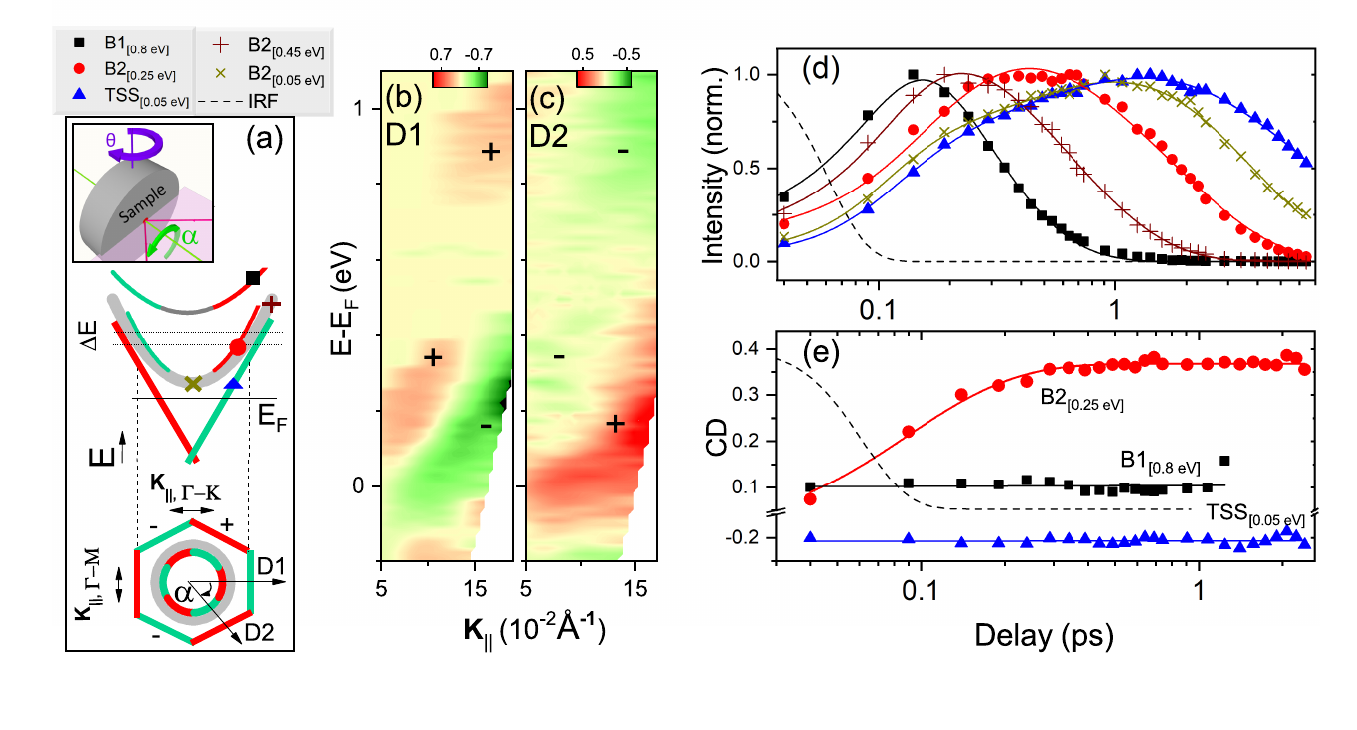}
\caption{ (a) Schematic representation of the out-of-plane component of spin-OAM texture \cite{sanchez2017,jung2011}. Red (green) color indicates the outward (inward) spin direction. The colored symbols mark relevant energy-momentum frame whose dynamics is shown in panels (d)-(e). Inset displays the rotation angles $\alpha$ and $\theta$. Panels (b) and (c) show the CD-ARPES map of the sample oriented along D1 and D2 at 250~fs delay (D1 and D2 are depicted in panel (a)). Panel (d) shows the normalized electronic dynamics of the corresponding symbols in (a). Solid lines represent best exponential fits. IRF is the instrumental response function, corresponding to the cross-correlation of pump and probe pulses. Panel (e) shows the evolution of CD for B1, B2 and TSS states extracted from areas shown in (a). Solid lines represent linear or exponential fits.}
\label{Fig2}
\end{figure*}

We also measured CD-ARPES at several azimuthal angles $\alpha$ (see Fig.~\ref{Fig2}a and Supplementary Information~\cite{SI}). Previous static spin-resolved ARPES and CD-ARPES measurements reported a strong hexagonal warping of the Dirac-cone in Bi$_2$Te$_3$ \cite{fu2009}. The distortion of the Dirac cone is amplified at larger $\mathbf{k}_{||}$ and eventually creates a snowflake-shape. The warped Dirac cone leads to an out-of-plane spin component which follows a three-fold symmetry and is maximum along $\bar{\Gamma}\bar{K}$ \cite{sanchez2014anisotropic}, as schematically reported in Fig.~\ref{Fig2}a. The sign-reversal of CD following a $\sin(3 \alpha)$ law was observed experimentally \cite{wang2011,jung2011} and supported theoretically \cite{mirhosseini2012}. We rotated the sample by $\alpha = 60^\circ$ (see Fig.~\ref{Fig2}a), checked the orientation by LEED and repeated the CD-trARPES measurements along the direction D2 (azimuthal rotation of $60^\circ$ relative to D1). Figure~\ref{Fig2}c shows the results at 250 fs delay and should be compared side-by-side with the D1 map (Fig.~\ref{Fig2}b): the sign reversal is clear. Additional measurements along other directions show consistent results according to the $ \sin (3\alpha)$ factor and the absence of CD along $\bar{\Gamma}\bar{M}$ or when the mirror plane of the crystal matches the incidence plane (see Supplementary Information \cite{SI}). We observed a similar behaviour not only for TSSs but also for B1 and B2. This observation indicates the three-fold symmetry of the CD signal of corresponding SR states, SR1 and SR2, since unpolarized BBs do not contribute to the CD signal of B1 and B2. The CD of SRs are anti-correlated with the one of the TSS, that has been predicted theoretically \cite{nurmamat2013} and our investigation confirmed it experimentally. We deduced the CD of SRs and TSS schematically shown in Fig.~\ref{Fig2}a. In the following, we elucidate the effect of photoexcitation on the total CD.  By excluding the effect of the pump pulse on the CD signal, since the other factors influencing the matrix elements remain constant, we are able to resolve the photoemitted electrons from different bands.

\begin{figure*}
\centering
\includegraphics[width=0.835\textwidth]{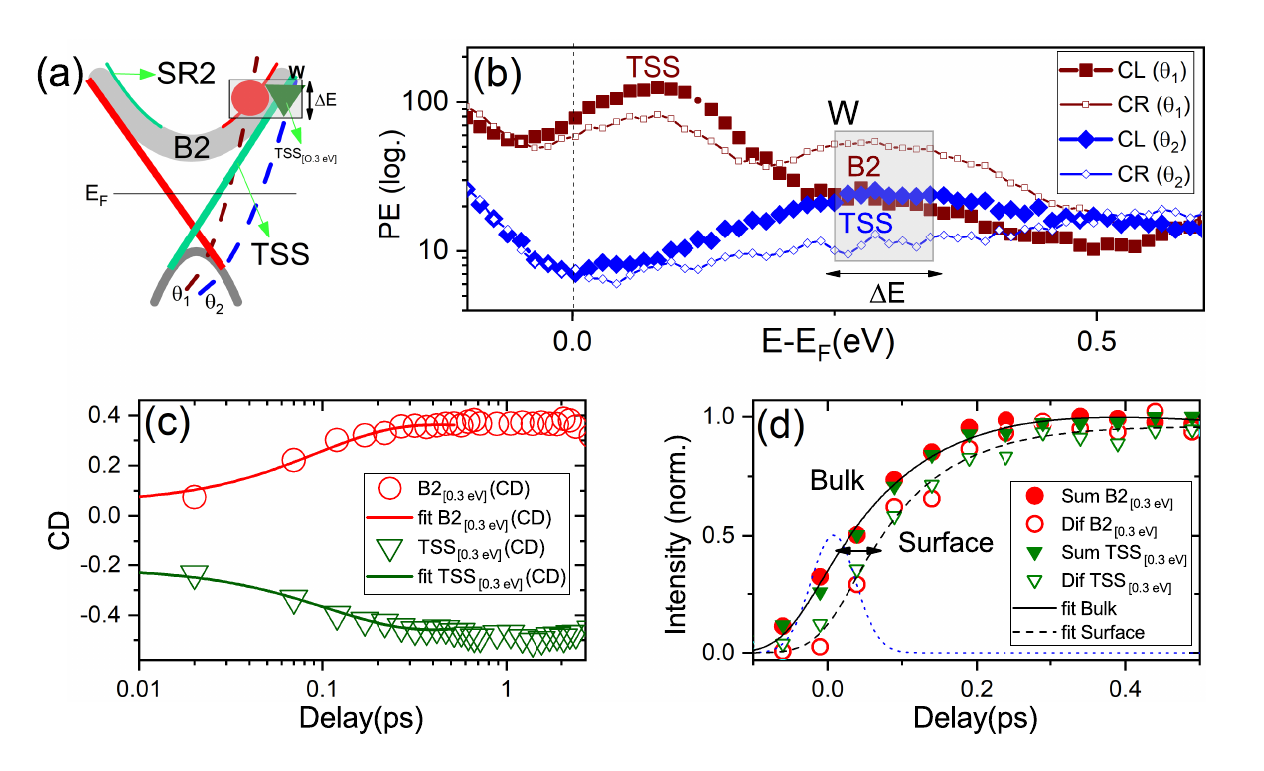}
\caption{Panel (a) schematically represents two angles of $\theta_{1}$ (brown) and $\theta_{2}$ (blue)  where the data of (b)-(d) were measured. The black dashed window (W) shows the important energy-momentum region where B2$_{[0.3~\rm{eV}]}$ (red circle) and TSS$_{[0.3~\rm{eV}]}$ (green triangle) are in the same energy level (see $\Delta$E in Fig.~\ref{Fig2}a ). Panel (b) shows the logarithmic EDCs of $\theta_{1}$ (brown) and $\theta_{2}$ (blue) taken with circular right (thin) and circular left (thick) probe polarization. (c) The time dependent CD of B2$_{[0.3~\rm{eV}]}$ and TSS$_{[0.3~\rm{eV}]}$. Solid lines are corresponding exponential fits. (d) The dynamics of sum circular light (S) and difference circular light (D). Solid line is corresponding fit to S and dashed line to D. the dotted Gaussian profile represents the instrumental response function.}
\label{Fig3}
\end{figure*}

We will introduce the following notation: X$_{[En]}$ represents the band X at a chosen binding energy En. The electronic dynamics of B1$_{[0.8~\rm{eV}]}$, B2$_{[0.25~\rm{eV}]}$ and TSSs$_{[0.05 ~\rm{eV}]}$ are extracted from the specified points of the band structure shown in Fig.~\ref{Fig2}a marked by squares, circles and triangles, respectively. The symbols "$+$" and "$\times$" indicates the dynamics in the upper and lower parts of B2 (B2$_{[0.45~\rm{eV}]}$ and B2$_{[0.05~\rm{eV}]}$). Since all bands are symmetric with respect to the $\bar{\Gamma}$ point, we extract the electronic dynamics of each band by referring to the energy of states, i.e. B1$_{[0.8~\rm{eV}]}$ is the B1 electronic states at $0.8~\rm{eV}$ above the Fermi level (see Supplementary Information~\cite{SI}). Figure~\ref{Fig2}d reports the corresponding normalized electronic dynamics. The pump pulse fills states at higher energies first. Hot electrons relax to lower states by scattering processes. The phenomenological fittings (exponential growth and decay convoluted with a Gaussian instrumental response function) show a fast sub 200 fs rise-time for all curves with binding energies $> 0.05~\rm{eV}$. To fit the rise-time of TSS$_{[0.05~\rm{eV}]}$ and B2$_{[0.05~\rm{eV}]}$, we use an additional exponential component with larger time constants of $0.72\pm0.1$~ps and $0.9\pm0.2$~ps, respectively. This behaviour can be explained by intra-band scattering of electrons to the states with lower energy, resulting in a a build-up at the bottom of the bands. This accumulation of electrons is attributed to the modest electron-phonon coupling \cite{tamtogl2017,sobota2014,luo2013snap} and to the lack of states$/$scattering channels for electron and hole recombination in an intrinsic TI~\cite{sterzi2017bulk,zhu2015,hedayat2018surface,sanchez2017}. Despite the similar rise-times of TSS$_{[0.05~\rm{eV}]}$ and B2$_{[0.05~\rm{eV}]}$, the decay occurs on significantly different time scales of $7.4\pm0.6$~ps and $3.10\pm0.26$~ps, respectively. We note that B2$_{[0.05~\rm{eV}]}$ is spin unpolarized and consequently has more scattering channels for relaxation with respect to TSS$_{[0.05~\rm{eV}]}$.

Figure~\ref{Fig2}e shows the time evolution of CD for B1$_{[0.8~\rm{eV}]}$, B2$_{[0.25~\rm{eV}]}$ and TSSs$_{[0.05 ~\rm{eV}]}$. While the CD signals of B1$_{[0.8~\rm{eV}]}$ and TTS$_{[0.05~\rm{eV}]}$ is essentially unaffected by the pump, the CD of B2$_{[0.25~\rm{eV}]}$ shows a clear rise-time. A stationary CD is consistent with the electronic dynamics of a spin-polarized band, since only electrons with the proper spin orientation can occupy these states, resulting in a constant dichroic signal, regardless of their number.  Note that the experimental geometry, the symmetry of initial and final states, the probe polarization and photon energy remain constant during the time evolution. We explored various factors which can influence the time-dependent CD induced by a linearly polarized pump pulse in order to determine the most plausible one. We first repeated the experiment with s-polarized pump beam and the results confirmed similar CD dynamics of B2$_{[0.25~\rm{eV}]}$ (see Supplementary Information~\cite{SI}). Therefore, the pump induced modification of matrix elements is excluded. In addition, the effect cannot be attributed to any change of the electronic population in the B2$_{[0.25~\rm{eV}]}$. This can be ruled out considering that, after reaching its maximum value within 600 fs from photoexcitation, the CD signal remains constant despite the almost complete loss of spectral weight (compare red circles in Fig.~\ref{Fig2}d and ~\ref{Fig2}e at large delays). Another possibility is a transient pump-induced spin polarization in B2$_{[0.25~\rm{eV}]}$, as reported in Refs.~\cite{hsieh2011,cacho2015,sanchez2016}. However, such a photoinduced effect should be pump-polarization dependent and also appear in B1$_{[0.8~\rm{eV}]}$, where the direct optical population is stronger. Therefore, we can exclude such effect. One other scenario is the accumulation of spin-polarized electrons in unpolarized BB due to spin-dependent decay channels. Only electrons with specific spin orientation can decay from B1 to TTS. The opposite spins accumulate in B2$_{[0.25~\rm{eV}]}$ which rises from B1 decay. However, owing to the depolarization and scattering channels of BBs, a long-lived spin population in the unpolarized B2$_{[0.25~\rm{eV}]}$ states (see Fig.~\ref{Fig2}e, red circles at long delays) is improbable. The last hypothesis is that in the probed region, we detect the superposition of un-polarized B2$_{[0.25~\rm{eV}]}$ and polarized SR2$_{[0.25~\rm{eV}]}$ photoelectrons giving rise to the variations of CD.

\begin{figure*}
\centering
\includegraphics[width=0.9\textwidth]{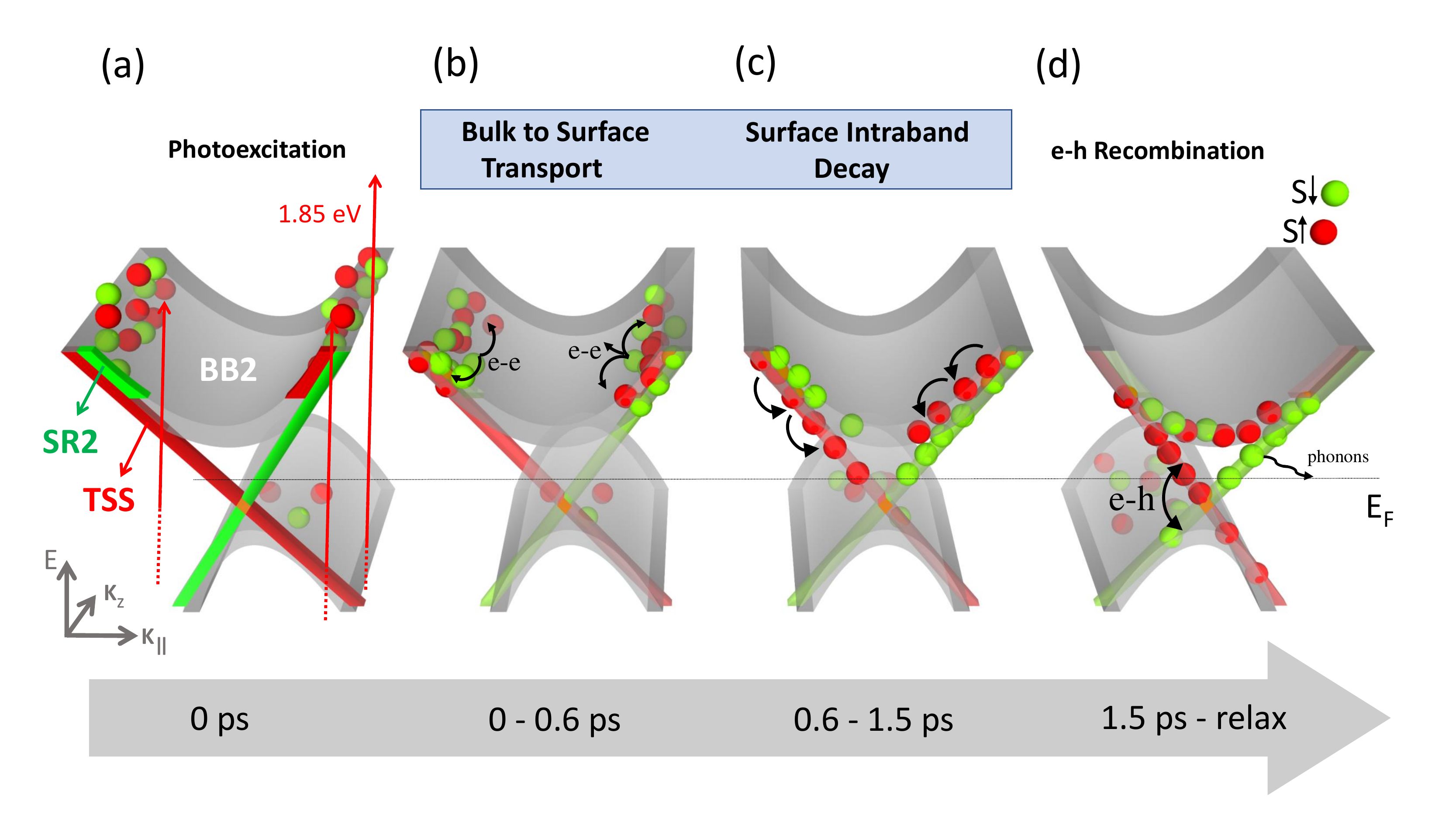}
\caption{The schematic representation of the electronic distribution after excitation of Bi$_2$Te$_3$ by a linearly polarized pulse. Each panel shows the dominant process. (a) First, upon the photoexcitation, the 1.85 eV pump promotes electrons above the Fermi level, hot electrons fill mainly the unoccupied BBs. (b) Electrons of BBs are rapidly scattered or transported to the surface. (c) Intraband scaterring process in TSSs and SRs. (d) Bottom of B2 and TSS relax by electron-hole recombination and energy transfer to phonons.}
\label{Fig4}
\end{figure*}

In order to confirm the last hypothesis, we analyzed CD at two specific photoemission angles $\theta_{1}\approx 9 ^\circ$ and $\theta_{2} \approx 16 ^\circ$ as depicted in Fig.~\ref{Fig3}a (see Fig.~\ref{Fig2}a for geometry of $\theta$ rotation). Note that B2 at $\theta_{1}$ and TSS at $\theta_{2}$ have the same binding energy $0.3~\rm{eV}$ (see the black dashed window, W, in Fig.~\ref{Fig3}a). Figure~\ref{Fig3}b shows the energy distribution curves (EDCs) at +250 fs delay, measured at emission angles $\theta_{1}$ and $\theta_{2}$ with both CR and CL probe polarization. We emphasize that, due to energy and $\mathbf{k}_{||}$ overlap, the PE signal from the surface localized SR2 states is mixed with the PE signal of B2 states. The unpolarized background (i.e. the helicity-independent spectral weight) is the contribution of B2: at angle $\theta_{1}$, it is mostly overlapped with SR2 and at $\theta_{2}$, it overlaps with TSS$_{[0.3~\rm{eV}]}$. TSS$_{[0.3~\rm{eV}]}$ is in contrast with TSS$_{[0.05~\rm{eV}]}$ of Figs.~\ref{Fig2}d-e which are taken at lower energies and are well separated from B2. Fig.~\ref{Fig3}b shows the TSS$_{[0.3~\rm{eV}]}$ and SR2$_{[0.3~\rm{eV}]}$ which display opposite CD (see the different intensities of CR and CL for each EDC in the black dashed window W). Figure~\ref{Fig3}c reports the dynamics of CD signal for B2$_{[0.3~\rm{eV}]}$ and TSS$_{[0.3~\rm{eV}]}$. Interestingly, the CD dynamics of TSS$_{[0.3~\rm{eV}]}$ is different with respect to the dynamics of  TSS$_{[0.05\rm{eV}]}$ reported in Fig.~\ref{Fig2}e. This is because TSS at higher energies (i.e. TSS$_{[0.3~\rm{eV}]}$) is strongly superimposed to B2$_{[0.3~\rm{eV}]}$. Presumably, the presence of B2$_{[0.3~\rm{eV}]}$ electrons in the PE signal causes the initial variations of the CD of TSS$_{[0.3~\rm{eV}]}$. To clarify this point, we compare the dynamics of CDs of B2$_{[0.3~\rm{eV}]}$ and TSS$_{[0.3~\rm{eV}]}$. The B2 electrons are present in both cases and we must detect a common dynamics. We fit CDs from  B2$_{[0.3~\rm{eV}]}$ and TSS$_{[0.3~\rm{eV}]}$ by an exponential function obtaining similar rise-times of $93 \pm 4$~fs, but opposite signs, the signature of a common underlying dynamics. In the energy region marked by box W, we attribute the common time-dependent signal in CDs to B2$_{[0.3~\rm{eV}]}$ unpolarized electrons since surface and bulk bands overlap. In fact, the constant CDs after about 600~fs show that the predominant PE signal comes from TSS$_{[0.3~\rm{eV}]}$ and SR2$_{[0.3~\rm{eV}]}$ at longer delays. Therefore, the relaxation dynamics is mostly a surface mechanism in the picosecond regime. 
Figure~\ref{Fig3}d compares the dynamics of B2$_{[0.3~\rm{eV}]}$ and TSS$_{[0.3~\rm{eV}]}$ when taking the sum of opposite helicities, S$=I_{CR}+ I_{CL}$, and the difference between opposite helicities D$=I_{CR}- I_{CL}$. "S" contains information about the dynamics of B2$_{[0.3~\rm{eV}]}$ (i.e. the unpolarized background). Fig.~\ref{Fig3}d shows that the dynamics of S is similar for B2$_{[0.3~\rm{eV}]}$ and TSS$_{[0.3~\rm{eV}]}$ (see solid symbols). This evidence suggests the presence of B2 electrons, a common spin-degenerate dynamics in both cases. Instead, D is mainly related to the spin-polarized SR2 and TSS. The 50 fs delay between S (B2$_{[0.3~\rm{eV}]}$) and D (SR2$_{[0.3~\rm{eV}]}$ or TSS$_{[0.3~\rm{eV}]}$) indicates that the photoexciation predominantly involves bulk states. Furthermore, we find exact match between the dynamics of SR2$_{[0.3~\rm{eV}]}$ and TSS$_{[0.3~\rm{eV}]}$ (open symbols). Figures~\ref{Fig3}c and ~\ref{Fig3}d show that the unpolarized B2 can be disentangled from SRs and TSSs on the femtosecond regime. Indeed, when electrons relax from bulk bands into the surface states, the effect of B2 in the probed region becomes less detectable. Our results suggest that in the measured energy-momentum window W, the spin-polarized surface electrons of TSS and SR2 appear with a delay after perturbation, then dwell at the surface for the entire relaxation process.

These findings allow us to differentiate the complex dynamics of TSSs, SRs and BBs. Although some recent studies focused on the spin decay behavior in TIs, a complete dynamical view of the electronic redistribution upon photoexcitation cannot be achieved without considering the role of TSSs, SRs and BBs. Experimentally, we observed a much faster decay of B2 states with respect to SRs and TSSs which is attributed to the 3D unpolarized nature of bulk bands with larger number of available decay channels. This agrees with previous studies in which the electron-electron scattering rate of the bulk bands of TIs has been suggested to be an order of magnitude larger than that of the TSSs \cite{sanchez2016}. The electronic transition between states with opposite spins requires a spin flip event. As a result, some transitions are hindered. This is in line with our observation that the B2 and TSS remain almost isolated from each other apart from the zone center, where the decay channel from the bottom of B2 becomes effective. Thus, our analysis provide a more comprehensive picture of the spin-based relaxation mechanisms. In this context, Fig.~\ref{Fig4} sketches the time-dependent electronic population and relaxation dynamics of Bi$_2$Te$_3$. At zero delay (Fig.~\ref{Fig4}a) the pump pulse promotes electrons from the occupied valence bands to empty bulk states. Then, electrons of unpolarized bulk bands migrate to surface (-resonant) bands according to the available spin states of each one. The oposite spin texture of the TSS and SR favors inter-band scattering for one spin direction and forbids it for the opposite direction. Consequently, intra-band electron decay is enhanced with respect to inter-band relaxation and explains the lack of interaction between SR2 and TSS during the relaxation time (Fig.~\ref{Fig4}c). Our results demonstrate that in an intrinsic TI, Bi$_2$Te$_3$, the picosecond relaxation process is mainly due to surface (resonant) states with strongly limited scattering channels. The complete relaxation occurring at long delays is explained by the energy transfer from the bottom of the TSS to low energy phonons \cite{sobota2014,tamtogl2017,luo2013snap} and electron-hole recombination \cite{freyse2018impact} at the surface (Fig.~\ref{Fig4}d).
\section{Conclusions\label{Conclusions}}
In summary, we have investigated the ultrafast electronic dynamics of Bi$_2$Te$_3$ combining circular dichroism with trARPES.  The results showed that the excitation with 1.85~eV pump photon energy takes place in bulk states, with a consequent ultrafast transport and redistribution of electrons in the surface. SR acts as a reservoir to accommodate the electrons with spins opposite to those in the TSS and plays a key role in re-establishing equilibrium. Distinguishing between bulk and surface dynamics has a fundamental importance for TI-based spintronic devices. In this context, our technique directly maps the unoccupied band structure and extracts the femtosecond light-induced dynamics resolving different unoccupied bands. Our study opens a new route to study the time-dependent electronic behavior in the bulk and surface of other TIs. We believe our results will trigger future theoretical and experimental studies on the SRs and their contribution to the electronic relaxation of a wide class of TI compounds.

\section{Methods\label{experiment}}

Single crystals of Bi$_{2}$Te$_{3}$ were grown using the self-flux method. The stoichiometric mixture of high purity elements was heated to 1000$^\circ$C for 12 hours and  then gradually cooled down to 500$^\circ$C over 100 hours before reaching room temperature. The samples were cleaved {\it in situ} and measured in ultrahigh vacuum conditions at pressure $<5\times 10^{-10}$~mbar and at room temperature.

TrARPES experiments were conducted using a Yb-based regeneratively amplified laser system with repetition rate of 100~kHz. The pump (1.85~eV photon energy, 30~fs pulse duration and p-polarization) and probe (6.05~eV, 65~fs) pulses, impinging on the sample at an incidence angle of about 45$^\circ$, were focused to spot sizes of about 100 and 50 $\mu$m, respectively \cite{boschini2014}. The time resolution of the experiment (width of the instrumental response function) is about 80~fs, and the pump fluence $\sim  300~\mu$ J/cm$^2$. Photoemission (PE) spectra were recorded using a time-of-flight (ToF) analyzer with an energy resolution of about 50 meV and angular acceptance of $\sim $0.8$^\circ$ \cite{carpene2009versatile}. The angle-resolved maps were acquired by rotating the sample's normal with respect to the analyzer axis by 3$^\circ$ steps. CD data were obtained using a $\lambda$/4 waveplate on the probe beam to generate circularly polarized light. The sample orientation was checked \textit{in-situ} by LEED.

\def\bibsection{\section*{\refname}} 
\bibliography{bibliography}

\section*{Acknowledgments\label{ack}}
E. C. and H. H. acknowledge support from PRIN $2017 - 2017$BZPKSZ$002$. G. C. acknowledges support by the European Union Horizon 2020 Programme under Grant Agreement No. $881603$ Graphene Core $3$.\\ 

\section*{Author contributions}
E. C., G. C. and C. D. conceived and coordinated the project. H. H. and D. B. performed the CD and trARPES experiments. H. Y., C. C. and X. J. Z synthesized Bi$_{2}$Te$_{3}$ single crystals. H. H. and E.C. analyzed the data and wrote the manuscript with contribution from all authors. All authors reviewed the manuscript and discussed the results.

\section*{Competing interests}
Te authors declare no competing interests.

\section*{Additional information}
\textbf{Supplementary information} is available for this paper at https://doi.org/xxx \\
\textbf{Correspondence} and requests for materials should be addressed to E. C.

\end{document}